\documentclass[useAMS,usenatbib]{mn2e}
\usepackage{graphicx}
\title{Scaling relations of the slightly self-interacting cold dark matter 
in galaxies and clusters}
\author[Chan]{M.~H.~Chan
\thanks{mhchan@phy.cuhk.edu.hk}\\
Department of Physics and Institute of Theoretical Physics, The Chinese
University of Hong Kong\\
Shatin, New Territories, Hong Kong, China}

\begin{document}

\date{Accepted XXXX. Received XXXX}

\pagerange{\pageref{firstpage}--\pageref{lastpage}} \pubyear{XXXX}

\maketitle

\label{firstpage}

\begin{abstract}
Recent observations in galaxies and clusters indicate dark matter 
density profiles exhibit core-like structures which contradict to the 
numerical simulation results of collisionless cold dark matter. The idea 
of self-interacting cold dark matter (SICDM) has been invoked to solve the 
discrepancies between the observations and numerical simulations. In this 
article, I derive some important scaling relations in 
galaxies and clusters by using the long-range SICDM model. These scaling 
relations give good agreements with the empirical fittings from 
observational data in galaxies and clusters if the dark matter particles 
are only slightly self-interacting. Also, there may exist a 
universal critical optical depth $\tau_c$ that characterizes 
the core-like structures. These 
results generally support the idea of SICDM to tackle the long-lasting 
dark matter problem.
\end{abstract} 

\begin{keywords}
Dark matter, galaxies, clusters
\end{keywords}

\section{Introduction}
The nature of dark matter remains a fundamental problem in 
astrophysics and cosmology. The rotation curves of galaxies and the mass 
profile probed by the hot 
gas in clusters indicate the existence of dark matter. It is commonly 
believed that dark matter is collisionless and becomes 
non-relativistic after decoupling. Therefore, they are regarded as cold 
dark matter (CDM). The CDM model can provide excellent fits on large scale 
structure observations such as Ly$\alpha$ spectrum \citep{Croft,Spergel}, 
2dF Galaxy 
Redshift Survey \citep{Peacock} and Cosmic Microwave Background 
\citep{Spergel2}. 

However, on the cluster and galactic scales, the CDM model shows 
discrepancies 
from observations. N-body simulations based on the CDM theory predict that 
the density profile of the collisionless dark matter halo should be 
singular at the center ($\rho \sim r^{\alpha}$). \citet{Navarro} first 
obtained $\alpha=-1$ (the NFW profile). Later, different values of 
$\alpha$ ranging from $-0.75$ to $-1.5$ were obtained 
\citep{Moore,Klypin,Taylor,Colin,Diemand}. Recently, 
high resolution numerical simulation indicates $\alpha=-0.8$ for $r 
\approx 120$ pc and $\alpha=-1.4$ for $r \approx 2$ kpc \citep{Stadel}. 
Nevertheless, observations show us core-like structures instead of 
singular 
density profile in many clusters and galaxies. For example, 
H$\alpha$ observations indicate cores present in over a hundred of 
disk galaxies and dark matter dominated galaxies 
\citep{Salucci,Borriello}. Later, \citet{deBlok} get a mildly cuspy slope 
$\alpha=-0.2 \pm 0.2$ based on modelling the presence of realistic 
observational effects. In cluster scale, 
observational data from gravitational lensing also show that cores exist 
in some clusters \citep{Tyson,Newman}. In particular, \citet{Sand} get 
$\alpha=-0.45 \pm 0.2$ by the combination of gravitational lensing and 
dynamical data of clusters MS2137-23 and Abell 383. Clearly, observations 
do not support the numerical small-scale predictions by the CDM model. 
This discrepancy is known as the core-cusp problem \citep{deBlok2}. 

In addition, 
computer simulations predict that there should exist thousands of small 
dark halos or dwarf galaxies in the Local Group if the dark matter 
particles are collisionless \citep{Cho}. However, observations of the 
Local Group only 
reveal less than one hundred galaxies \citep{Spergel}. Such discrepancy is 
known as the missing satellites problem \citep{Cho}. 

Many theories have been invoked to solve the core-cusp problem and the 
missing satellites problem. One of the most spectacular idea is that the 
dark matter is not cold. The existence of keV sterile neutrinos, as a 
candidate 
of warm dark matter (WDM), has been proposed to solve the discrepancies 
\citep{Xue}. However, recent observations tend to reject the keV sterile 
neutrinos to 
be the major component of dark matter since the observational bound of 
sterile neutrino mass in Lyman-alpha forest contradicts to that in x-ray 
background \citep{Abazajian,Viel,Seljak}. Also, the WDM model alone cannot 
get a good 
agreement on the large scale power spectrum \citep{Spergel,Boyarsky}. The 
WDM model is likely 
to be ruled out in standard cosmology. Therefore, the success of the CDM 
model on large scales suggests that a modification of the dark matter 
properties may be the only approach to solve the discrepancies 
\citep{Spergel}. \citet{Spergel} proposed that the conflict of 
observations and simulations can be reconciled if the CDM particles are 
self-interacting. Later, \citet{Burkert} performed the numerical 
simulation 
of the self-interacting cold dark matter (SICDM) and showed that core-like 
structures can be produced. On the 
other hand, the analysis of the metallicity distributions of 
globular clusters indicates that the existence of the SICDM is able to 
solve the missing satellites problem \citep{Cote}. The earliest estimated 
range of the cross-section per unit mass of the self-interacting dark 
matter particle is
$\sigma/m=(0.45-450)$ cm$^2$ g$^{-1}$ \citep{Spergel}. This ratio has 
been estimated several times by some model dependent observations of 
clusters and galaxies and numerical simulations. For example, 
\citet{Randall} and \citet{Bradac} obtained $\sigma/m<0.7$ cm$^2$ 
g$^{-1}$ and $\sigma/m<4$ cm$^2$ g$^{-1}$ respectively by using the 
observational 
data from the clusters 1E 0657-56 and MACS J0025.4-1222. On the galactic 
scales, \citet{Ahn} and \citet{Koda} show that $\sigma/m \sim 100$ cm$^2$ 
g$^{-1}$ can explain the core-like structures. 

However, gravitational lensing and X-ray data indicate that the cores 
of clusters are dense and ellipsoidal where SICDM model predicts that to 
be shallow and spherical \citep{Loeb}. Therefore, the dark matter 
cross-section may either be smaller than expected or depend on velocity. 
Nevertheless, \citet{Peter} show that the discrepancies can still be 
solved even if the cross-section is velocity-independent. The latest 
numerical simulations with SICDM indicate that the cross-section per unit 
mass should be $\sigma/m \sim 0.01-0.1$ cm$^2$ g$^{-1}$ in order to 
produce the reported core sizes and central densities of galaxies and 
clusters \citep{Buckley,Rocha,Peter,Zavala}.

In this article, I will show 
in another way that the slightly long-range interaction of dark matter can 
naturally generate some model 
independent scaling relations in galaxies and clusters which 
agree with the observations. Lastly, I will comment on this small 
interaction of dark matter.

\section{Optical depth of the dark matter particles}
In SICDM model, the size of a core in a structure depends on the 
self-interacting rate of the dark matter particles. This rate is closely 
related to a physical quantity `optical depth of the dark matter 
particles' $\tau$. The optical depth for dark matter is defined as 
$\tau \equiv n 
\sigma d$, where $d$ is the distance travelled by a dark matter 
particle and $n$ is the mean number density of the dark matter 
particles. Therefore, the optical depth within the core radius $r_c$ is 
given by $\tau=n \sigma r_c$. The dark matter particles can be considered 
as collisonless if $\tau \approx 0$. \citet{Spergel} propose that $\tau 
\approx 1$ within the core, which corresponds to the `photosphere' of the 
dark matter. However, this optical depth is too large to match the 
observational data. Here, we assume that the size of the core is 
characterized by a critical optical depth $\tau_c$ such that
$n \sigma r_c=\tau_c$, where $0 \le \tau_c \le 1$. Since the core mass is 
given by $M_c=4 \pi mn r_c^3/3$, we have
\begin{equation}
n \sigma r_c= \frac{3M_c}{4 \pi r_c^2} \left( \frac{\sigma}{m} 
\right)=\tau_c.
\end{equation}
The above equation indicates a rough scaling relation $M_c \propto 
r_c^2$ if $\tau_c$ is a constant. This relation is 
generally consistent with the recent result in galaxies obtained by 
\citet{Gentile}: $M_c=72^{+42}_{-27}\pi r_c^2M_{\odot}$ pc$^{-2}$.
 
\section{The Scaling relations in clusters and galaxies}
\subsection{Baryonic Tully-Fisher relation}
The orbital speed in a galaxy is given by 
\begin{equation}
V=\sqrt{\frac{GM}{R}},
\end{equation}
where $M$ and $R$ are the total enclosed mass and radius of luminous 
matter respectively. From Eq.~(2), the observed flat rotation curves 
in most galaxies give $M/R \approx M_c/r_c$. By combining Eqs.~(1) and 
(2), we get
\begin{equation}
M_c= \left( \frac{3}{4 \pi \tau_c} \right) \left( \frac{\sigma}{m} 
\right)G^{-2}V^4.
\end{equation}
The density profile of the SICDM can be approximately given by the Burkert 
profile \citep{Burkert2,Rocha}:
\begin{equation}
\rho(r)=\frac{\rho_0r_c^3}{(r+r_c)(r^2+r_c^2)},
\end{equation}
where $\rho_0$ is the central density of dark matter. Therefore, the 
integrated mass profile is given by
\begin{equation}
M(r)=\int_0^r4 \pi r^2 \rho(r)dr= \pi \rho_cr_c^3f(r), 
\end{equation}
where $f(r)=\ln[(r^2+r_c^2)/r_c^2]+2 \ln[(r+r_c)/r_c]-2 \tan^{-1} 
(r/r_c)$. The size of luminous matter $R$ can be regarded as the 
radius $r_{max}$ where the rotation curve peaks in the simulations, ie. $R 
\approx r_{max}$. Since the 
numerical simulations indicate that $r_{max} \approx 3r_c$ \citep{Rocha}, 
by Eq.~(5), the integrated total mass to core mass ratio is about $M/M_c 
\approx 5$. 
Assume that the ratio of total baryonic mass to total 
mass is nearly a constant for all galaxies ($M_b/M \approx 
\Omega_b/ \Omega_m \approx 0.17$, where $\Omega_b$ and $\Omega_m$ are the 
cosmological 
density parameters of baryonic matter and total matter respectively), the 
total baryonic mass of a galaxy is
\begin{equation}
M_b= \left( \frac{15}{4 \pi \tau_c} \right) \left( \frac{\sigma}{m} 
\right) 
\left( \frac{\Omega_b}{\Omega_m} 
\right)G^{-2}V^4.
\end{equation}
If $\tau_c$ and $\sigma/m$ are constant for all galaxies, we have $M_b 
\propto V^4$. This scaling relation is indeed the baryonic 
Tully-Fisher relation \citep{Tully,McGaugh,McGaugh2}. Latest observations 
indicate $M_b=(47M_{\odot}$ km$^{-4}$ 
s$^{-4})$$V^4$ \citep{McGaugh2}. If $\sigma/m=0.1$ cm$^2$ g$^{-1}$, 
we get $\tau_c=0.005$. In fact, \citet{Mo} have already shown 
that the Tully-Fisher relation can be obtained by assuming a particular 
form of cored density profile. Here, I use another independent and 
simpler way to show that the Tully-Fisher relation is consistent with the 
SICDM scenario. 

Furthermore, from Eq.~(1), we 
have $\rho_0r_c=\tau_c(\sigma/m)^{-1}$, which would be a constant if 
$\tau$ and $\sigma/m$ are constants. Surprisingly, recent analysis 
indicates that $\rho_0r_c=141^{+82}_{-52}M_{\odot}$ pc$^{-2}$, which is a 
constant for a large sample of dwarf and late-type galaxies 
\citep{Gentile}. If $\tau=0.005$ and $\sigma/m=0.1$ cm$^2$ g$^{-1}$, we 
get $\rho_0r_c \approx 240M_{\odot}$ pc$^{-2}$, which is 
generally closed to the empirical fits from observations.

\subsection{Size-Temperature relation in clusters}
\citet{Reiprich} studied more than 100 clusters' hot gas profiles and 
probed the total mass of each cluster. The mass profile of a cluster can 
be approximately given by \citep{Reiprich}
\begin{equation}
M(r) \approx \frac{3 \beta kTr^3}{Gm_g(r^2+r_c^2)},
\end{equation}
where $\beta$ is the parameter ranging from $0.4-1.1$ in the King's 
$\beta$-model \citep{King}, $T$ is the hot gas temperature and $m_g$ is 
the mean mass of a hot gas particle. Here, we have 
used the fact that the hot gas profiles are nearly isothermal in most 
clusters \citep{Reiprich}. From Eq.~(7), the central density of the dark 
matter is given by
\begin{equation}
\rho_0= \frac{9 \beta kT}{4 \pi Gm_gr_c^2}.
\end{equation}
Since the central density of hot gas is just $10^{-26}$ g cm$^{-3}$ 
\citep{Mohr2}, which 
is much less than the total central density $10^{-23}$ g cm$^{-3}$, the 
effect of the baryons at the centre is ignored. By combining Eqs.~(1) and 
(8), we get
\begin{equation}
r_c \approx \left( \frac{9}{4 \pi \tau_c} \right) \left(\frac{\sigma}{m} 
\right) \left(\frac{\beta k}{Gm_g} \right)T.
\end{equation}
In fact, $r_c$ represents the core sizes of both total matter 
(dominated by dark matter) and baryonic matter \citep{Reiprich}. 
Therefore, the size of the hot gas in cluster can be characterized by 
$r_c$. If $\tau_c$ and $\sigma/m$ are constant for all clusters, we have 
a scaling relation $r_c \propto T$, which 
agrees with the empirical fits $R'=0.5(T/6~\rm keV)^{1.02}~\rm Mpc$ 
from observational data of some nearby clusters \citep{Mohr,Sanders3}, 
where $R'$ is the isophotal size of a cluster. For a 
$10^{15}M_{\odot}$ cluster, $r_c \approx 300$ kpc \citep{Rocha}, which is 
$\approx 
0.7R'$. If $\sigma/m=0.1$ cm$^2$ g$^{-1}$, by using Eq.~(9) and the mean 
$\beta$ for all clusters, we have $\tau_c=0.006$.

\subsection{Mass-Temperature relation in clusters}
Besides the Size-Temperature relation, we can also obtain a scaling 
relation of the total cluster mass and hot gas temperature. At large 
radii, the hot gas in clusters may not be isothermal. The total 
cluster mass will be closed to the Burkert mass profile in Eq.~(5): 
\begin{equation}
M \approx 7.9 \pi \rho_0r_c^3,
\end{equation}
where we have assumed that $R_{200} \approx 15r_c$ and $R_{200}$ is 
the radius when the mean total mass density equals to 200 times 
cosmological critical density. By putting 
Eqs.~(8) and (9) into the above equation and assuming $M_b/M \approx 
\Omega_b/\Omega_m$, we have
\begin{equation}
M_b \approx \left( \frac{40}{\pi \tau_c} \right) \left(\frac{\sigma}{m} 
\right) 
\left( \frac{\Omega_b}{\Omega_m} \right) \left( \frac{\beta k}{Gm_g} 
\right)^2T^2.
\end{equation}
Since the hot gas mass dominates the baryonic mass in most clusters, the 
total hot gas mass $M_g$ in a cluster is closed to the total 
baryonic mass 
$M_b$. This scaling relation $M_g \approx M_b \propto T^2$, again, agrees 
with the empirical fitting from clusters 
$M_g/10^{14}M_{\odot}=0.017(T/1~\rm keV)^2$ \citep{Mohr2,Sanders3}. By 
putting all the known numerical values and $\sigma/m=0.1$ cm$^2$ 
g$^{-1}$ into Eq.~(11), we get $\tau_c=0.002$.

\section{Discussion}
In this article, I show that the long-range interaction of CDM can 
naturally obtain some important scaling relations, including the 
baryonic Tully-Fisher relation for galaxies ($M_b \propto V^4$), the 
Size-Temperature relation ($r_c \propto T$) and Mass-Temperature relation 
($M_g \propto T^2$) in clusters. These scaling 
relations get remarkably good agreements with the empirical fits from 
observations. If the cross-section per unit mass is $\approx 0.1$ 
cm$^2$ g$^{-1}$, the characteristic critical optical depth $\tau_c \approx 
0.002-0.006$, which is the nearly the same values in different scaling 
relations for 
galaxies and clusters. Moreover, we can get $\rho_0r_c \approx 
240M_{\odot}$ pc$^{-2}$, which is a constant for all galaxies. This result 
is generally consistent with 
the recent analysis from the observations of dwarf and late-type galaxies 
\citep{Gentile}. It means only a slight dark matter interaction is 
enough for producing core-like structures. Therefore, when the central 
density is high enough such that $\tau=\tau_c$, a core would be produced. 
It may explain why some clusters do not exhibit core-like structures as 
their central densities are too low such that $\tau \leq \tau_c$ within a 
resolvable radius. 

In the past decade, it is believed that the dark matter cross-section is 
velocity-dependent \citep{Koda,Loeb}. Nevertheless, recent results in 
simulations show that it is possible to have a velocity-independent 
cross-section $\sigma/m \approx 0.1$ cm$^2$ g$^{-1}$ \citep{Rocha,Peter}. 
In this model, the scaling relations derived may support this idea and 
enable us to measure this small cross-section by using observational data. 
Although only a small window of constant cross-section is remained 
\citep{Zavala}, our results provide more evidences to support the SICDM 
scenario, which can successfully address the dark matter problem, 
core-cusp problem and the missing satellite problem.

To conclude, the derived scaling relations by using the SICDM scenario can 
get good agreements with observations. It generally supports the idea of 
the SICDM and the velocity-independent dark matter cross-section. More 
simulations and observations 
will be needed to confirm the existence of the universal critical optical 
depth $\tau_c$, which characterizes the core-like structures in galaxies 
and clusters.

\label{lastpage}

\end{document}